\documentclass{elsart}
\usepackage{graphicx} 
\begin{document}
\begin{frontmatter} 

\title{Scaling of local interface width of statistical growth models}
\author[UFF]{Anna Chame\corauthref{cor1} }  
\ead{achame@if.uff.br}
\corauth[cor1]{Fax number: (55) 21-2629-5887 }
\author[UFF]{F. D. A. Aar\~ao Reis}
\ead{reis@if.uff.br}           
\address[UFF]{Instituto de F\'\i sica, Universidade Federal Fluminense,\\
Avenida Litor\^anea s/n, 24210-340 Niter\'oi RJ, Brazil}
\date{\today}
\maketitle

\begin{abstract}
We discuss the methods to calculate the roughness exponent $\alpha$ and the 
dynamic exponent $z$ from the scaling properties of the  local roughness, which
is frequently used in the analysis of experimental data.
Through numerical simulations, we studied the Family, the restricted
solid-on-solid (RSOS), the Das Sarma-Tamborenea (DT) and the Wolf-Villain (WV)
models in one- and two dimensional substrates,  in order to compare different
methods to obtain those exponents. The scaling at small length scales do
not give reliable  estimates of $\alpha$, suggesting that the usual methods
to estimate that exponent from experimental data may provide misleading
conclusions concerning the universality classes of the growth processes. On the
other hand, we propose a more efficient method to  calculate the dynamic
exponent $z$, based on the scaling of characteristic correlation lengths,
which gives estimates in good agreement with the expected universality
classes and indicates   expected crossover behavior. Our results 
also provide
evidence of Edwards-Wilkinson asymptotic behavior for the DT and the WV models
in two-dimensional substrates. 
\end{abstract}
 

\end{frontmatter}
                   
\section{Introduction}
\label{intro}

The comparison of morphological features of thin films' surfaces
and of those of discrete or continuum growth models is of fundamental
importance to infer the basic mechanisms of the experimental growth
processes~\cite{frontiers,barabasi,Krug}. Statistical models usually
represent real systems' features by simple stochastic processes, neglecting
the details of the microscopic interactions, but still being able to reproduce
their large scale properties. Frequently, the interest is to classify
model systems and real surfaces into universality classes of interface growth.
At this point it is essential that the theoretical systems be
investigated in the same lines of the experimental work, i. e. by analyzing
the same physical quantities with standard methods.

In the study of interface growth models, one usually is interested in the
scaling properties of the global interface width. For a discrete deposition
model in a $d$-dimensional substrate of length $L$, the global width is
defined as
\begin{equation}
\xi(L,t) = {\left[ { \left< { { {1\over{L^d}}
\sum_i{ {\left( h_i - \overline{h}\right) }^2 } } } \right> } \right] }^{1/2} ,
\label{defxi}
\end{equation}
where $h_i$ is the height of column $i$ at time $t$, the bar in
$\overline{h}$ denotes a spatial average and the angular brackets denote a
configurational average. For short times (growth regime), $\xi$ is expected to
scale as
\begin{equation}
\xi \sim t^{\beta} , 
\label{defbeta}
\end{equation}
and for long times (steady state) it saturates as
\begin{equation}
\xi_{sat} \left( L\right) \equiv \xi\left( L,t\to\infty \right) \sim
L^{\alpha} .
\label{defalpha}
\end{equation}
The dynamical exponent is $z=\alpha/\beta$.

In numerical studies, the exponent $\beta$ is measured in the growth regime
of very large substrates. The exponents $\alpha$ and $z$
are obtained from data in the steady states or approaching this long-time
regime, in which the heights of the deposits ($\overline{h}\sim t\sim L^z$
or larger, with $z\geq 1$) significantly exceed their lateral lengths ($L$).

On the other hand, in experiments and in some theoretical works
(analytical and numerical) one is interested in the scaling properties of {\it
local} surface fluctuations during the growth regime, when $\overline{h}\ll
L$ and, consequently, the effects of finite lateral sizes of the substrate are 
negligible. In these conditions, height fluctuations inside small windows
(boxes) over a very large surface are measured.  This is achieved by
calculating the height-height correlation function $G\left( r,t\right) \equiv
\left<{\overline{ {\left( h_{i+r}-h_i\right) }^2} } \right>$ or the local
interface width
\begin{equation}
w(r,t) \equiv {\langle{ {\left< {\left[ h_i- {\left< h\right>}_r \right]}^2
\right>}_r}\rangle}^{1/2} ,
\label{deflw}
\end{equation}
where ${\left< \dots\right>}_r$ denotes a spatial average over windows of size
$r$. Usually, these windows are square boxes of side $r$ scanning the
surface of the deposit.
$\sqrt{G\left( r,t\right)}$ and $w(r,t)$ have the same
scaling properties. In systems with normal scaling (in opposition to anomalous
scaling), the local width scales as
\begin{equation}
w(r,t) \sim t^{\beta}  f\left( r/t^{1/z} \right) ,
\label{fvlw}
\end{equation}
where $f$ is a scaling function. It is expected that
\begin{equation}
f(x)\to x^{\alpha} , x\ll 1
\label{fxll1}
\end{equation}
and
\begin{equation}
f(x)\to const , x\gg 1 .
\label{fxgg1}
\end{equation}
In systems with anomalous scaling, Eq. (\ref{fvlw}) is still valid, but $f$
scales with $\alpha_{local}<\alpha$ for small $x$ (small $r$)~\cite{lopez}
(it has been shown~\cite{Tang} that $\alpha_{local} \leq 1 $).

Experimentally, the roughness exponent $\alpha$ is usually obtained from the
scaling of the local width or the height-height correlation
function at small length scales (Eq. \ref{fxll1}).  
Some techniques have been developed to 
characterize a surface  by using a few  images of varying 
scan sizes or even only one image subdivided into windows of a given 
size. The scaling of the local roughness (Eqs. \ref{fvlw} and \ref{fxll1}) is
then used to estimate the exponent $\alpha$ \cite{Krim,Rao}.  However, one of
the main problems for the calculation of $\alpha$ is the narrow range in which
$\log{w}$ increases approximately linearly with $\log{r}$. Sometimes that
range does not exceed one order of magnitude. In some theoretical works,
similar procedure was also adopted, although it is more common the calculation
of $G\left( r,t\right) $ or $w\left( r,t\right)$ in the steady states, i. e.
for very long deposition times~\cite{Rubia}. Even then,  the problem of a
restricted scaling region (Eq. \ref{fxll1}) is still present; see e. g. Ref.
\protect\cite{Siegert}. 

The first aim of this work is to study the scaling properties of the {\it
local} interface width of several limited-mobility growth models in order to
propose methods to calculate the scaling exponents
from data in the growth regimes, i. e. for very large system sizes
and relatively small times. The advantages or disadvantages of each method may
guide the lines of investigation of the universality classes of real growth
processes. One of our conclusions is that the accuracy of the estimate of the
exponent $\alpha$ is very poor when it is calculated with a method that
parallels the one used in experimental works, i. e. based on the scaling 
relation (\ref{fxll1}) for small length scales. For some theoretical models
with weak corrections to scaling, this problem may be partially overcome with
another method, but this method is not suitable for analyzing experimental 
data due to their typical error bars. On the other hand, we present a method to
calculate the exponent $z$ from the local roughness scaling and discuss its
advantages with application to some discrete models. We will show that, in
experimental works where the interest is to search for the universality
classes of the growth processes from local width data, the best choice 
seems to be the calculation of the exponent $z$ with that technique.

The models we will consider here are the random deposition with
surface relaxation of Family~\cite{family}, the  restricted  solid-on-solid
(RSOS) model of Kim and Kosterlitz~\cite{kk} and  the molecular  beam epitaxy
models of Das Sarma and Tamborenea (DT model)~\cite{dt} and  of Wolf and  
Villain (WV model)~\cite{wv}, in $1+1$ and $2+1$ dimensions.
From the theoretical point of view, the 
advantage of the method proposed here is that the computational cost is very
much reduced when compared to a study in which one has to wait until the 
stationary regime is attained. Also, our results are particularly 
relevant to elucidate recent questions~\cite{chatraphorn,toroczkai}
on the universality classes of the WV and the DT models in $2+1$ dimensions.

The rest of this work is organized as follows.
In Sec. 2, we will briefly review the growth rules of the discrete models
studied here, the continuum equations representing the expected universality
classes and the simulation procedure. In Sec. 3, we will discuss the
methods to estimate the roughness exponent in the growth
regime. In Sec. 4, we will present the method to calculate the dynamical
exponent. In Sec. 5 we will present our results for the DT and the WV 
models. In Sec. 6 we summarize our results and present our conclusions. 

\section{Models, universality classes and simulation procedure}

We will simulate four limited mobility growth models whose
stochastic aggregation rules are illustrated in Figs. 1a-d.

\begin{figure}[ht]
\centering
\includegraphics[clip,width=0.80\textwidth, 
height=0.40\textheight,angle=0]{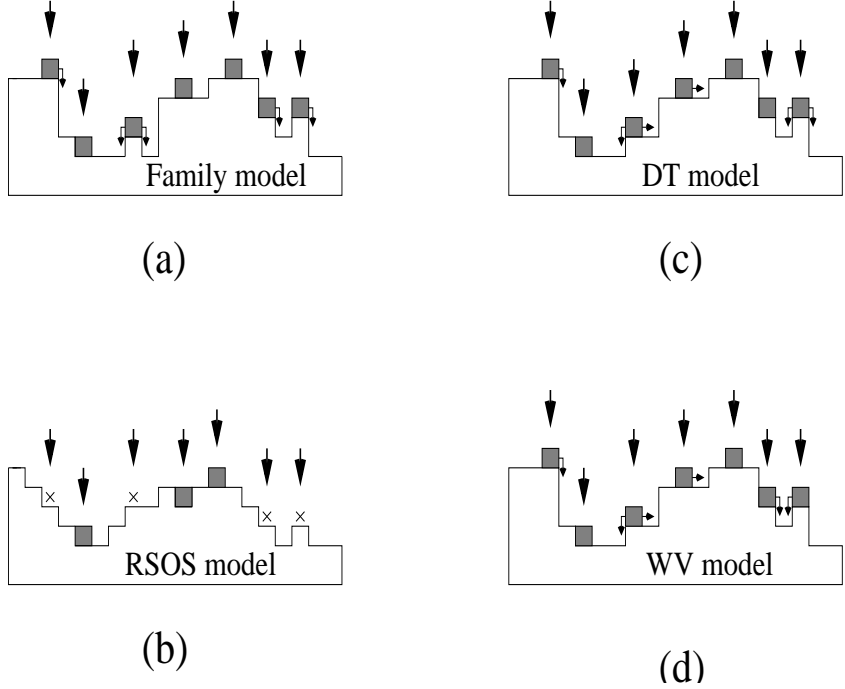}
\caption{ (a) The aggregation rules of the Family model, in which the
relaxation of incident particles to their sticking positions is indicated by
arrows. The incident particle in the middle has equal probabilities to
stick at any one of the neighboring columns.(b) The aggregation rules of the $RSOS$
model, where the crosses indicate the rejected deposition attempts.
c) The aggregation rules for the DT model, where the deposited particle  seeks
only to have one nearest neighbor and (d) for the WV model, where the
particle will choose the site with the largest number of neighbors. }
\label{fig1}                        
\end{figure}

 In the Family model~\cite{family} (Fig. 1a), a column of 
the deposit is randomly chosen and, if no neighboring column has a smaller 
height than the
column of incidence, the particle sticks at the top of this one. Otherwise, it
sticks at the top of the column with the smallest height among the neighbors.
If two or more neighbors have the same minimum height, the sticking position is
randomly chosen among them.

In the continuum limit, the Family model is expected to belong to the
Edwards-Wilkinson (EW) universality class~\cite{ew}, i. e. its
scaling properties are the same obtained from the linear EW equation
\begin{equation}
{{\partial h}\over{\partial t}} = \nu{\nabla}^2 h + \eta (\vec{x},t) .
\label{ew}
\end{equation}
Here, $h$ is the height at the position $\vec{x}$ at time $t$, $\nu$
represents a surface tension and
$\eta$ is a Gaussian noise~\cite{barabasi,kpz} with zero mean and 
variance
$\langle \eta\left(\vec{x},t\right) \eta (\vec{x'},t' ) \rangle =
D\delta^d (\vec{x}-\vec{x'} ) \delta\left( t-t'\right)$.
The EW equation can be exactly solved (\cite{ew} - see also Ref.
\protect\cite{Krug}), giving $\alpha =1/2$ in $d=1$ 
and $\alpha=0$ ($\xi^2\sim \ln{L}$) in $d=2$, while $z=2$ in all dimensions.

In the $RSOS$ model~\cite{kk} (Fig. 1b), the incident particle may stick at the
top of the column of incidence if the differences of heights between 
the incidence column and each of the  neighboring columns
do not exceed ${\Delta h}_{MAX} = 1$. Otherwise, the
aggregation attempt is rejected. Due to the dependence of the local growth rate
on the local height gradient, the RSOS model is asymptotically
represented by the Kardar-Parisi-Zhang equation~\cite{kpz}
\begin{equation}
{{\partial h}\over{\partial t}} = \nu{\nabla}^2 h + {\lambda\over 2}
{\left( \nabla h\right) }^2 + \eta (\vec{x},t) ,
\label{kpz}
\end{equation}
where $\lambda$ represents the excess velocity. The exponents of the KPZ class
in $d=1$ are $\alpha=1/2$ and $z=3/2$~\cite{barabasi,kpz}, and in $d=2$ they
are $\alpha\approx 0.4$ and $z\approx 1.6$~\cite{barabasi}.

The DT and WV models were originally proposed to represent molecular-beam
epitaxy.

In the DT model (Fig. 1c), a column $i$ of the deposit is randomly chosen
and the incident particle sticks at the top of that column if it has one or
more lateral neighbors at that position. Otherwise, the neighboring columns
(at the right and the left sides in $d=1$) of column $i$ are 
tested. If the
top position of only one of these columns has, at least, one lateral neighbor,
then the incident particle aggregates at that point. If no neighboring column
satisfies this condition, then the particle sticks at the top of column $i$.
Finally, if two or more neighboring columns have, at least, one lateral
neighbor, then one of them is randomly chosen.

Theoretical approaches~\cite{predota,huang} predict that the
$1+1$-dimensional DT model is described, in  the continuum limit, by the
Villain-Lai-Das Sarma (VLDS) growth  equation~\cite{villain,laidassarma}
\begin{equation}
{{\partial h}\over{\partial t}} =
\nu_4{\nabla}^4 h + \lambda_{22} {\nabla}^2 {\left( \nabla h\right) }^2 + \eta
(\vec{x},t) . \label{vlds}
\end{equation}
where $\nu_4$ and $\lambda_{22}$ are constants and $\eta$ is a Gaussian noise.
Eq. (\ref{vlds}) gives exponents $\alpha = 1$, $\beta =1/3$ and
$z=3$ in $d=1$ and gives $\alpha = 2/3$, $\beta =1/5$ and
$z=10/3$ in $d=2$. The crossover of the exponents of the DT model to those of
the VLDS theory in $d=1$ was discussed in recent works~\cite{punyindu,brunoc},
but simulations using noise-reduction schemes~\cite{chatraphorn,toroczkai,CTS}  
provided estimates of exponent $\beta$ in $d=2$ which disagree with the VLDS
theory  and  found that the asymptotic behavior of the DT model in $d=2$ 
is in the EW class.

In the WV model (Fig. 1d), the growth rules are slightly different from those
of the DT model. After choosing the column of incidence $i$, the incident
particle aggregates at the top of the column with the largest number of
lateral neighbors. If there is a tie between column $i$ and one or more
neighboring columns, then the particle aggregates at column $i$. Otherwise, in
the case of a tie between neighboring columns, one of them is randomly chosen.

In the continuum limit, the WV model in $d=1$ is expected to belong to the EW
class. Indeed, many works have already analyzed the long crossover to the
asymptotic exponents in that case~\cite{smilauer,kotrla,park,Siegert,brunoc}.
Krug et al \cite{KPS} and Siegert\cite{Siegert} observed a crossover to the
EW class in $d=2$, but the recent works of Das Sarma and
collaborators~\cite{chatraphorn,toroczkai,CTS} suggested the unstable 
(mounded morphology) universality class in that case.

Here, the $1+1$-dimensional models will be simulated in lattices of lengths
$L=262144$ and periodic boundary conditions, which is suitable to represent an
infinitely large substrate. The maximum simulation time (measured in number of
deposition attempts per site) for the Family and the RSOS models is
$t_{max}=8\times{10}^3$. The maximum simulation time
for the DT and the WV models is much larger, $t_{max}=256\times{10}^3$, in
order to analyze the crossover to the asymptotic exponents of these
controversial problems.

The calculation of the local interface width is done with one-dimensional
boxes of length $r$ in the range $4\leq r\leq 65536$.  For each size $r$, 
the box glides through the lattice (in such a way that  one of its  
extremities visits successively each site of the lattice)  
and for each box position the local roughness is calculated,  
giving a  contribution  to the average $<  >_r$.  

In $d=2$, lattices of lengths $L=2048$ are considered, and maximum simulation
times ranged from $t_{max}=8\times{10}^3$ (RSOS and Family models) to
$t_{max}=3.2\times{10}^4$ (DT and WV models). Local widths are calculated 
within gliding square boxes of lengths ranging from $r=2$ to $r=400$ in most 
cases.

\section{Calculation of roughness exponents}
\label{alpha}

In Fig. 2 we show the local width $w$ as a function of the box size $r$ for the
Family and the RSOS models in $d=1$, at $t= 8\times{10}^3$. The dashed line
has slope equal to the exponent $\alpha =0.5$ of the EW and the KPZ theories
in $d=1$. The scaling form (\ref{fxll1}) for small $r$ predicts linear
behavior in that log-log plot. However, for the RSOS model, the deviations are
clearly visible in Fig. 2 if two decades of the variable $r$ are considered.
For the Family model the deviations appear within a narrower range of $r$.

\begin{figure}[h]
\centering
\includegraphics[clip,width=0.80\textwidth,
height=0.40\textheight,angle=0]{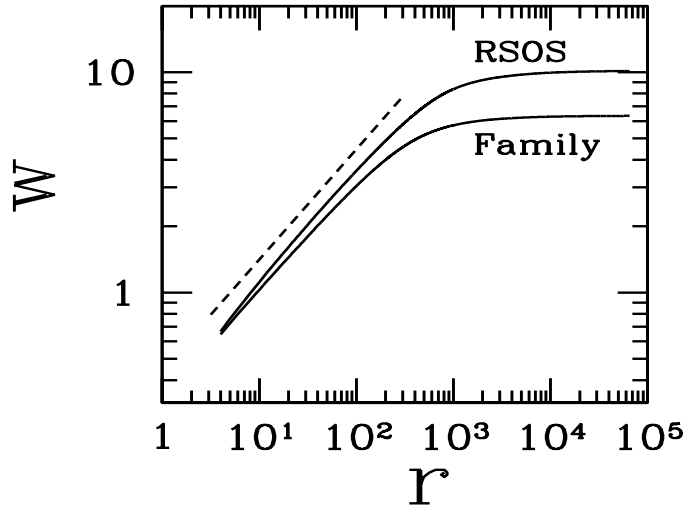}    
\caption{  The local width $w$ as a function of the box size $r$ for the
$1+1$ dimensional Family model and for the $1+1$-dimensional RSOS model,
at $t= 8\times{10}^3$. The dashed line has slope $0.5$.
}
\label{fig2}
\end{figure}  

We conclude that linear fits of $\log{w}\times\log{r}$ plots are not
reliable to provide estimates of roughness exponents which indicate 
the true universality class of the process. In order to make this point
clearer, we calculated the consecutive slopes of the $\log{w}\times\log{r}$
plots,
\begin{equation}
\alpha(r,t) \equiv {{\Delta\log{w}}\over{\Delta\log{r}}} .
\label{alphaeff}
\end{equation}
The effective exponents $\alpha(r,t)$ are shown in Fig. 3a as a function of $r$
for the RSOS model and in Fig. 3b for the Family model, in both cases for three
different deposition times. They show inflection points at
$\alpha\left( r,t\right) \approx 0.5$, which will ultimately turn
into plateaus with $\alpha\left( r,t\right)$ equal to the asymptotic $\alpha$.
However, the deposition times will have to increase many orders of
magnitude and, consequently, the deposit will not have the features of a thin
structure anymore.

\begin{figure}[h]
\centering
\includegraphics[clip,width=0.80\textwidth,
height=0.40\textheight,angle=0]{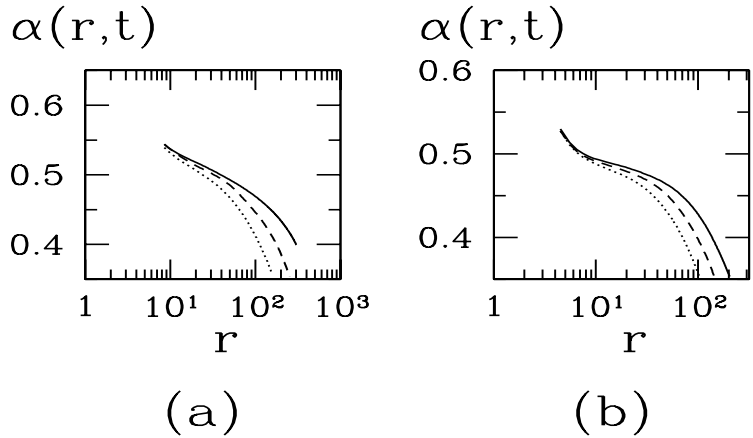}   
\caption{Effective exponents $\alpha(r,t)$ as a
function of $r$ for (a) the $1+1$-dimensional RSOS model  and for (b) the
$1+1$-dimensional Family model, in
both cases for three different deposition times: $t=2000$ (dotted line),
$t=4000$ (dashed line), $t=8000$ (solid line).
 }
\label{fig3}
\end{figure}

This kind of problem was already observed in the
scaling of the correlation function of the $2+1$ dimensional WV
model~\cite{Siegert}. However, while the WV model presents a long crossover to
an asymptotic behavior (to be discussed later), the Family and the RSOS models 
present very nice scaling properties when the global width $\xi$ is analyzed. 
In other words, simple extrapolation methods provide accurate estimates of the
exponent $\alpha$ from saturation widths $\xi_{sat}$ in small lattices.
Thus, the above problems lead to the
conclusion that the local width scaling in the growth regime is not suitable
for calculating an exponent $\alpha$ which reliably indicates the
class of the growth process. This is particularly important in experiments
where the local roughness (or the correlation function) scaling in the growth
regime is analyzed, because the slope of a linear fit of an
arbitrarily chosen region of the $\log{w}\times\log{r}$ plot may lead to a
value of $\alpha$ which incorrectly identifies the universality class.

Another  problem  has been previously
found~\cite{Rubia} in the scaling of the local  width, for which  there are  
corrections due to  effects of  finite spatial resolution, which would be
relevant only for  systems with $\alpha_{local}>1/2$. However, this is 
not the  case of the Family and the RSOS models.

In order to partially overcome the problems above, one possibility is the
analysis of the slopes of the $\alpha(r,t)\times \log{r}$ plots, whose minima
seem to indicate the asymptotic value of $\alpha$. This is indeed achieved in
the plots of Figs. 4a and 4b, where we show ${{d\alpha{\left(
r,t\right)}}\over{d\log{r}}}$ (the local curvature of the
$\log{w}\times\log{r}$ plot) as a function of $\alpha(r,t)$ for the RSOS and
the Family models, respectively. For both models, the minimum of
${{d\alpha{\left( r,t\right)}}\over{d\log{r}}}$ decreases in time, which
indicates an increasingly better fit of the $\log{w}\times\log{r}$ data to a
straight line, and the corresponding $\alpha\left( r,t\right)$ converges to
the asymptotic $\alpha$.

\begin{figure}[h] 
\centering
\includegraphics[clip,width=0.80\textwidth,
height=0.40\textheight,angle=0]{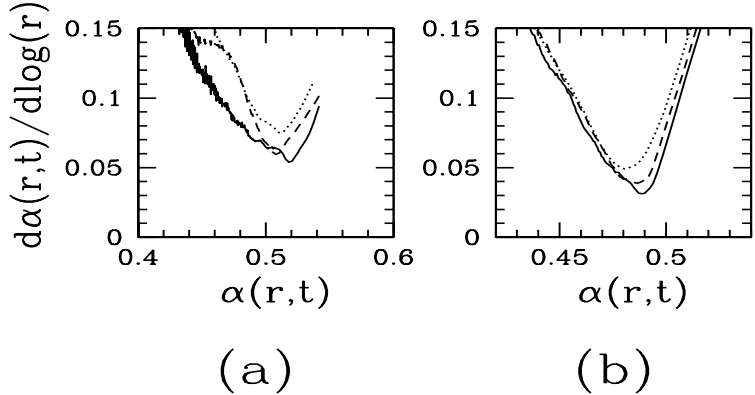}   
\caption{Slopes of the $\alpha(r,t)\times \log{r}$ plots as a function of
$\alpha(r,t)$ for: (a) $1+1$-dimensional RSOS model; (b) $1+1$-dimensional Family model.
In both cases, data for three different deposition
times are shown: $t=2000$ (dotted line), $t=4000$ (dashed line), $t=8000$
(solid line). The position of the minima in the figures tend to the asymptotic
$\alpha$.}
\label{fig4}
\end{figure}  

It is important to stress that this procedure is usually not suitable for the
analysis of experimental data due to the difficulties to calculate second
derivatives with reasonable accuracy. Thus, its interest is restricted to
theoretical work. Moreover, our results for the DT and WV models, to be
presented in Sec. 5, will show that this procedure  does not work properly 
for models with significant crossover effects.

The scenario is not very different in $d=2$. The effective exponents
$\alpha(r,t)$ behave similarly to those in Figs. 3a for the RSOS model
and the minima of ${{d\alpha{\left( r,t\right)}}\over{d\log{r}}}$ suggest
$\alpha\approx 0.4$. For the Family model, an approximately logarithmic
growth of the squared local width (giving $\alpha =0$) is obtained, but also
in a restricted range of $r$.

\section{Calculation of dynamical exponents}
\label{z}

In order to estimate the exponent $z$, our first step is to calculate a
characteristic length $r_c$ which is proportional to the correlation length at
a given time $t$. This is obtained by defining $r_c$ as
\begin{equation}
w\left(r_c,t\right) = k\xi\left( t\right) ,
\label{defrc}
\end{equation}
where $\xi\left( t\right)$ is the global width at time $t$ and $k$ is a
constant. From Eqs. (\ref{defbeta}) and (\ref{fvlw}), it is expected that
\begin{equation}
r_c \sim t^{1/z} .
\label{scalingrc}
\end{equation}
Geometrically, $r_c$ is the abscissa of the $w\times r$ plot at which $w$
attains a fixed fraction $k$ of its saturation value ($\xi$).
This method is inspired on those previously used to estimate 
crossover times and dynamical exponents from the global 
width~\cite{liu,fabio,albano}.

Typically, the values of $k$ are chosen so that $r_c>10$ and $r_c\ll L$, where
$L$ is the total length of the substrate (the latter condition have to be more
flexible in $2+1$ dimensions). Here, we will generally consider values of $k$
between $0.6$ and $0.9$.

For fixed $k$, effective exponents $z_n(t)$ are defined as
\begin{equation}
z_n(t) \equiv  { {2 \ln{n}}\over 
{\ln\left[ r_c\left( nt\right) / r_c\left( t/n\right) \right] } } ,
\label{zeff}
\end{equation}
which converge to $z$ when $t\to\infty$. Here, the values
$n=2$ and $n=4$ will be considered in Eq. (\ref{zeff}).

\begin{figure}[h]
\centering
\includegraphics[clip,width=0.80\textwidth,
height=0.40\textheight,angle=0]{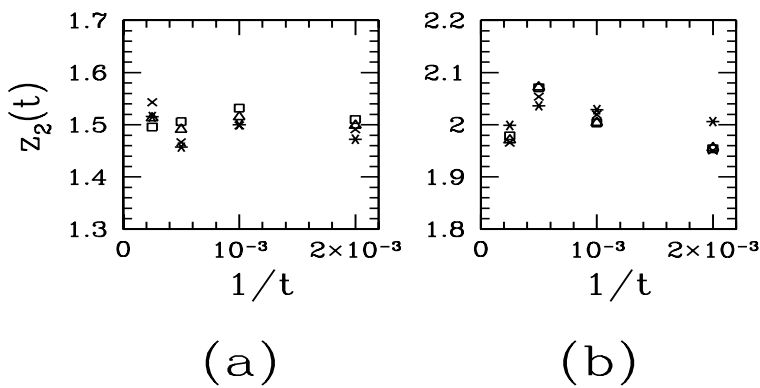}
\caption{Estimates of the effective exponents $z_2$ for: (a) 
$1+1$-dimensional  RSOS  model; (b) $1+1$-dimensional Family model.  For 
both models,
the  estimates for $r_c$ are obtained using
$k=0.6$ (squares), $k=0.7$ (triangles), $k=0.8$ (crosses)  and $k=0.9$
(asterisks).   }
\label{fig5}
\end{figure}

In Figs. 5a and 5b we show $z_2(t)\times 1/t$ for the RSOS and the Family
models in $d=1$. We notice that $z_2(t)$ oscillates around the expected
asymptotic values, $z=3/2$ (RSOS model) and $z=2$ (Family model), with
differences typically smaller than $10\%$, even using data from short
deposition times. It contrasts to the behavior of the effective roughness
exponents shown in Figs. 3a and 3b, which suggests that estimating the
dynamical exponent from the local widths is a better method to infer the
universality class of the process. Also note that there is no systematic
deviation of the data for different values of $k$, which is an important test
of the reliability of the method.

Before presenting results in $d=2$, we recall that the Family model (EW
class) shows logarithmic scaling in that case. Thus, the procedure to calculate
the characteristic length $r_c$ is different. The solution of the EW
equation~\cite{forrest,Krug} suggests the scaling form
\begin{equation}
w^2(r,t) = A\ln{\left[  g\left( r/t^{1/z} \right) t  \right]  } ,
\label{fvlwlog}
\end{equation}
where $A$ is a constant ($z=2$ in Eq. \ref{fvlwlog}). The saturation value
of the local width, for $r\to\infty$, is the global width
$\xi^2\left( t\right) =A\ln{\left( Bt\right)}$, where
$B=\lim_{x\to\infty}{g\left( x\right)}$. The characteristic length $r_c$ is
then defined so that
\begin{equation}
w^2{\left(r_c,t\right)} = \xi^2{\left( t\right)} -C ,
\label{defrclog}
\end{equation}
where $C$ is a positive constant. Consequently, $r_c$ is expected to scale as
Eq. (\ref{scalingrc}) with $z=2$ or, equivalently, the effective exponents
defined in Eq. (\ref{zeff}) should converge to $z=2$.

\begin{figure}[h]
\centering
\includegraphics[clip,width=0.80\textwidth,
height=0.40\textheight,angle=0]{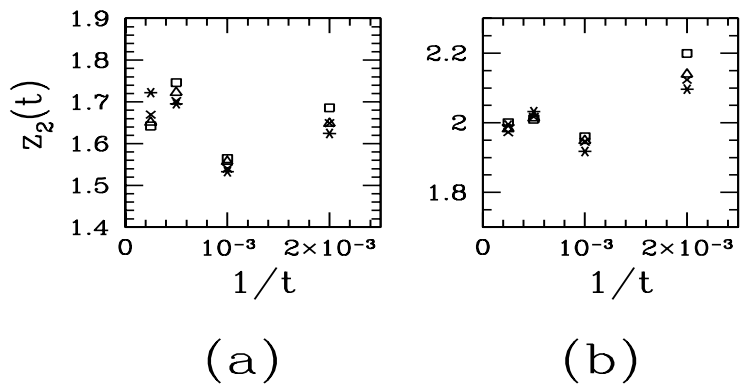}
\caption{Estimates of the effective exponents $z_2$ for: (a)
$2+1$-dimensional RSOS  model, assuming
power-law scaling for the local width, with $r_c$ obtained  using 
$k=0.6$
(squares), $k=0.7$ (triangles),  $k=0.8$ (crosses)  and $k=0.9$
(asterisks);
(b)  $2+1$-dimensional Family model assuming logarithm
scaling for the local width, with $r_c$ calculated using  $C=0.1$   
(asterisks), $C=0.15$ (crosses), $C=0.2$ (triangles) and $C=0.25$
(squares).
}
\label{fig6}
\end{figure}

In Fig. 6a we show $z_2(t)\times 1/t$ for the $2+1$-dimensional RSOS
model, with $r_c$ calculated from Eq. (\ref{defrc}) using $k$ from $0.6$ to
$0.9$. In Fig. 6b we show $z_2(t)\times 1/t$ for the $2+1$-dimensional Family
model, with $r_c$ calculated from Eq. (\ref{defrclog}) using $C=0.25$, $C=0.2$,
$C=0.15$ and $C=0.1$. The choice of the values of the constant $C$ obeys the
same criteria adopted for choosing $k$. In both cases, the effective exponents
oscillate around the expected asymptotic values, $z\approx 1.6$ for the
KPZ class and $z=2$ for the EW class. Again, the analysis of effective
dynamical exponents is superior to the analysis of effective roughness
exponents (except for the possibility of analyzing ${{d\alpha{\left(
r,t\right)}}\over{d\log{r}}}$ plots, but this is certainly limited to
theoretical work).

\section{Results for the DT and the WV models}

First we applied the method to estimate $\alpha$ (Sec. 3) to the DT and the
WV models in $d=1$. Although the deposition times were large (up to
$2.56\times{10}^5$ monolayers), the estimated exponents $\alpha$ were still
very far from the asymptotic values. For the DT model, the anomalous scaling
was theoretically predicted, with local roughness exponent $\alpha_{loc} =
8/11\approx 0.73$~\cite{lopez}. Our estimate $\alpha\approx 0.7$ is in good
agreement with that value. For the WV model, $\alpha\approx 0.75$ was
obtained, which is significantly higher than the EW value $\alpha =0.5$.
However, this discrepancy is expected because a very slow crossover to the
asymptotic behavior was already observed by several
authors~\cite{Siegert,Ryu,smilauer}.

In $d=2$, the scenario is the same. For the DT model, we obtained
$\alpha\approx 0.35$, which is not consistent with the VLDS value (well
established only in $1+1$ dimensions), but agrees with  a local roughness
exponent $\alpha = 0.3$ obtained from a  study~\cite{SP} of the anomalous
multiscaling of this model, which is recognized as a transient effect.  For the
WV model, $\alpha\approx 0.65$, which is distant both from the EW value (also
well established only in $1+1$ dimensions) and from the value $\alpha=1$
suggested by Das Sarma and co-workers~\cite{chatraphorn,toroczkai,CTS}.

Now we turn to the calculation of exponent $z$ (Sec. 4) of those models.

The results in $d=1$ give evidence of long crossovers but no reliable
extrapolation can be performed, as can be seen in Figs. 7a and 7b, which show
$z_4(t)\times 1/t^{1/2}$ for the DT and the WV models, respectively. The value
$n=4$ was used in Eq. (\ref{zeff}) because the differences in the estimates of
$r_c$ for consecutive times ($t$ and $2t$) was very small, which is a
consequence of the small values of $1/z$ (see Eq. \ref{defrc}). The abscissa
$1/t^{1/2}$ instead of $1/t$ was chosen to avoid superposition of data points
for large $t$. Although the data in Figs. 7a and 7b do not show a clear
convergence to the asymptotic values $z=3$ (DT model) and $z=2$ (WV model), we
note that the effective exponents clearly diverge from the value $z=4$ of the
fourth order {\it linear} growth theory (Eq. \ref{vlds} with
$\lambda_{22}=0$), which was suggested to represent their universality classes
in the original works~\cite{dt,wv}.

\begin{figure}[h]
\centering
\includegraphics[clip,width=0.80\textwidth,
height=0.40\textheight,angle=0]{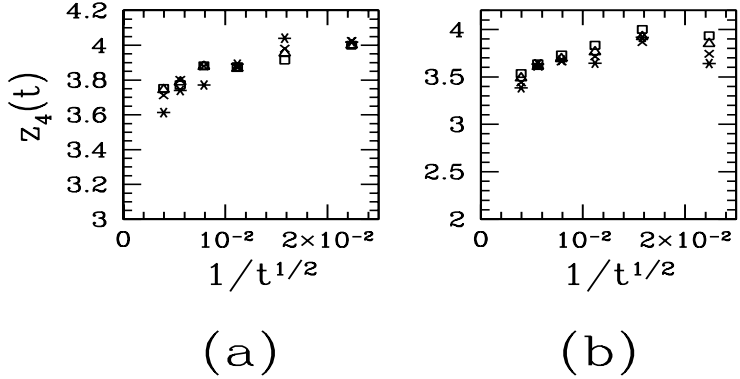}
\caption{Estimates of the effective exponents $z_4$ for: (a)
$1+1$-dimensional DT  model
(expected asymptotic exponent $z= 3$); (b) $1+1$- dimensional WV
model
(expected asymptotic exponent   $z= 2$).  For both models, the
estimates of $r_c$ are obtained using $k=0.6$ (squares), $k=0.7$ 
(triangles),$k=0.8$  (crosses)  and  $k=0.9$ (asterisks).
}
\label{fig7}
\end{figure}

\begin{figure}[h]
\centering
\includegraphics[clip,width=0.80\textwidth,
height=0.40\textheight,angle=0]{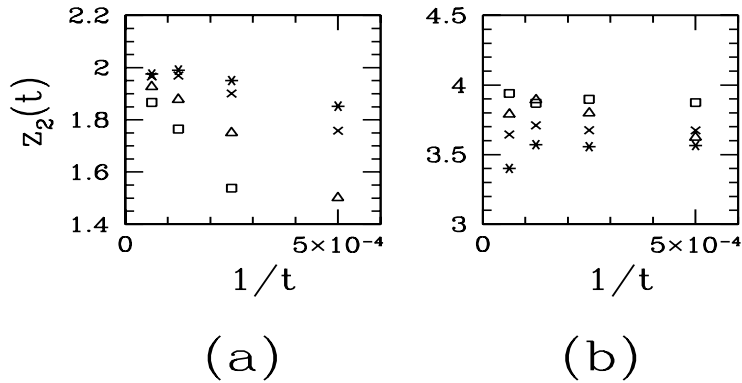}
\caption{Estimates of the effective exponents $z_2$ for the 
$2+1$-dimensional WV model: (a) assuming logarithmic scaling for the
local width, $r_c$ was obtained with $C=4$ (squares),
$C=3$ (triangles),  $C=2$ (crosses)  and $C=1.5$ (asterisks); (b)
assuming power law
scaling for the local width, $r_c$ was calculated using $k=0.6$ 
(squares),   $k=0.7$ (triangles),  $k=0.8$ (times)  and $k=0.9$ 
(asterisks).
}
\label{fig8}
\end{figure}

Now we consider separately those models in $d=2$.

In Fig. 8a, we show for the WV model the effective exponents $z_2(t)$ as a 
function of $1/t$,
obtained from the characteristic lengths $r_c$ calculated using Eq.
(\ref{defrclog}), which is suitable for logarithmic scaling of $w$. The
corresponding plot based on the assumption of power law scaling for the local
width (Eq. \ref{defrc}) is shown in Fig. 8b. The data in Fig. 8a clearly
converge to $z\approx 2$ as $t\to\infty$, suggesting that the WV model is also
in the EW class in $d=2$. Notice the consistence of the results for four
different values of $C$ in Eq. (\ref{defrclog}). On the other hand, with the
assumption of power law scaling for the local width, the effective exponents
for different $k$ (Eq. \ref{defrc}) tend to spread as $t$ increases (Fig. 8b).

The evidence of an asymptotic EW behavior for the WV
model in $d=2$ reinforces the conclusion of Siegert~\cite{Siegert}, who
observed a crossover in the scaling of the structure factor. On the other
hand, it is in contradiction with the suggested unstable (mounded morphology)
universality class for this model~\cite{chatraphorn,toroczkai,CTS}. At this
point, it is important to stress that this mounded morphology was observed in
simulations with noise reduction schemes, while here and in the paper of
Siegert the original WV model was considered.             

\begin{figure}[h]
\centering
\includegraphics[clip,width=0.80\textwidth,
height=0.40\textheight,angle=0]{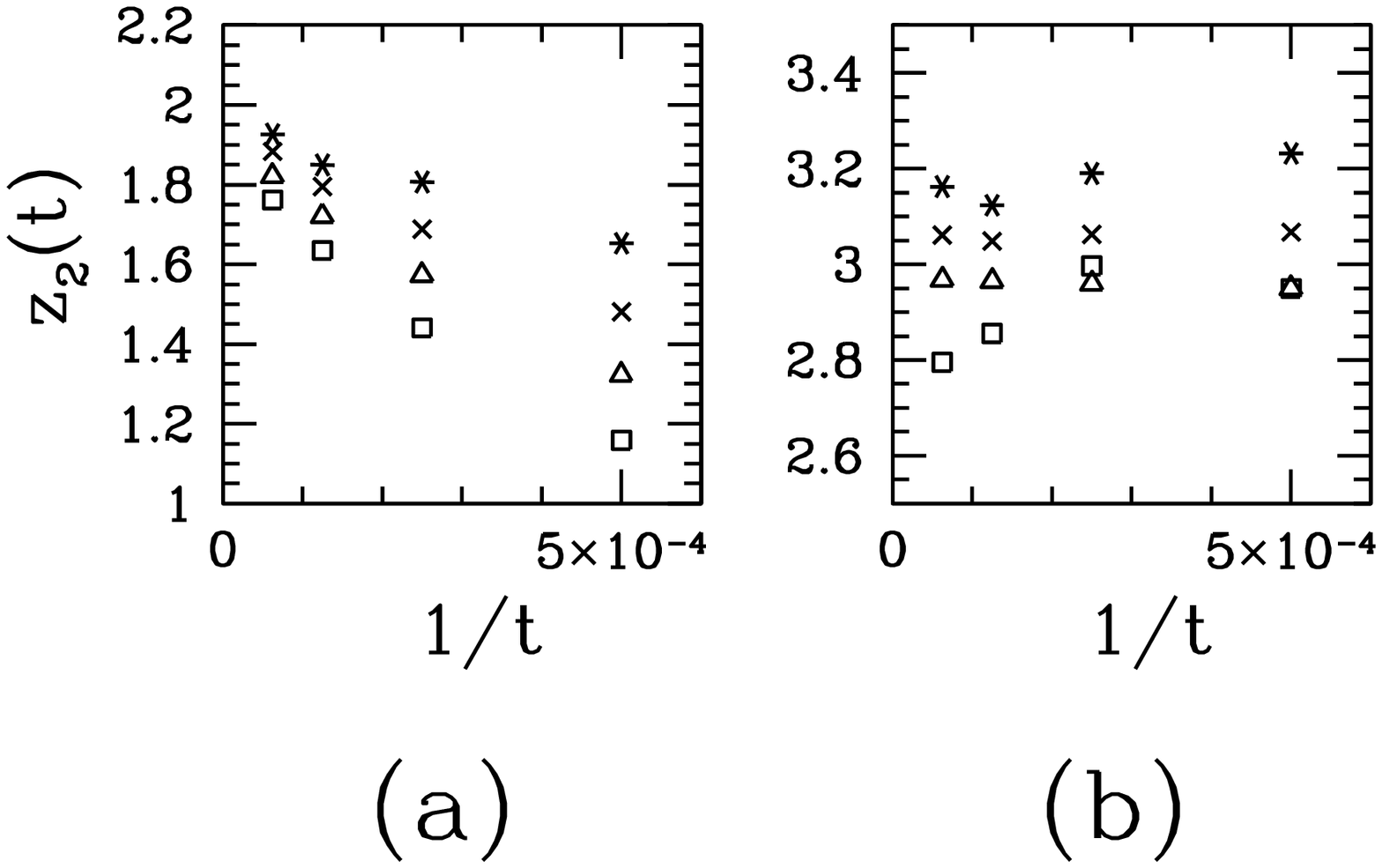}
\caption{Estimates of the effective exponents $z_2$ for the 
$2+1$-dimensional DT model: (a) assuming logarithmic scaling for the
local width, $r_c$ was obtained with $C=6$ (squares),
$C=5$ (triangles),  $C=4$ (crosses)  and $C=3$ (asterisks); (b)
assuming power law
scaling for the local width, $r_c$ was calculated using $k=0.6$ 
(squares),
$k=0.7$ (triangles),  $k=0.8$ (times)  and $k=0.9$ (asterisks).
}
\label{fig9}
\end{figure}

The same analysis was performed with the data for the DT model. In Fig. 9a 
we show $z_2(t)\times 1/t$ obtained with the assumption of logarithmic scaling
for calculating $r_c$ (Eq. \ref{defrclog}), and in Fig. 9b we show
$z_2(t)\times 1/t$ obtained with the assumption of power law scaling (Eq.
\ref{defrc}). The results in Fig. 9a strongly suggest that $z=2$
asymptotically. On the other hand, all data in Fig. 9b are smaller than
the value $z=10/3=3.333\dots$ of the  VLDS theory in $d=2$, and there is no
sign that those data will increase for larger deposition times. Instead, the
effective exponents for $k=0.6$ show a decreasing trend as $t$ increases.
Consequently, our data also suggests that the DT model is in the EW class in
$d=2$.

A comparison of the cases $d=1$ and $d=2$ is essential
at this point. First, in $d=1$ the power law scaling is justified by the
fact that the values of $z_4(t)$ obtained with different values of
$k$ were nearly the same. However, no extrapolation of $z_4(t)$ leads to an
asymptotic exponent consistent with the known classes of interface growth.
Thus, the best that can be said from our data is that there is a crossover
from the fourth order linear behavior ($z=4$) to a different class. On the other
hand, in $d=2$, different effective exponents were obtained with fixed $t$ and
different $k$ or $C$, but only the logarithmic scaling hypothesis led to the
same asymptotic behavior for different $C$. Simple extrapolations are possible
and give $z\approx 2$, thus indicating the asymptotic EW class.

Another important point on the usefulness of the method to 
calculate $z$ is that, if we want to determine whether the system presents  
logarithmic scaling or not, we can simply test power-law  and 
logarithmic behaviors and analyze the convergence (or divergence) of the
estimates of $z$ as time increases, for different choices of a single parameter
($C$ for logarithmic behavior, $k$ for  power-law behavior). The fact that 
there is not an optimal  value for $k$ or $C$ give us increasing confidence
in the  extrapolated value, since for the correct choice of the
dynamic scaling relation the asymptotic $z$ must not depend on these
parameters.

\section{Conclusion}
\label{conclusion}

We studied the scaling of the local interface width $w$ of several limited
mobility growth models, focusing on the methods to estimate the scaling
exponents. The methods to calculate the roughness exponent from the scaling of
$w$ at small length scales do not give reliable estimates since the 
intervals of window sizes in which that scaling is valid are small.
On the other hand, the method proposed to calculate the dynamical exponent
provides effective exponents in agreement with the expected universality
classes for models with weak scaling corrections and reflects the expected
crossover behavior for models such as DT and WV in $1+1$ dimensions. The
difficulties to measure the roughness exponents may be partially overcome in
theoretical studies by improving the analysis of the $log{w}\times\log{r}$
plots, but it only works for models with very weak scaling corrections. This
analysis leads to the conclusion that the calculation of the exponent $z$ from
experimentally measured local widths is more adequate than the calculation 
of $\alpha$ to infer the universality class of the growth process. 

We believe that the method based on the local interface width  would be 
equivalent to the method based on the height-height correlation 
function. For instance, our estimate of (effective) $\alpha$ for the
$2+1$-dimensional DT  model is consistent with the estimate obtained from
the scaling of the height-height correlation function by Das Sarma and
Punyindu~\cite{SP}, which was recognized as a transient regime. 
We agree with the observation of Siegert~\cite{Siegert} that the exponents
$\alpha $ and $\beta$ obtained from the scaling of the height-height
correlation function (or the local width) are not so reliable as those
obtained from other methods, such as the structure factor. Thus, our
proposal is to calculate the dynamical exponent $z$ whenever it is necessary to
deal with local roughness data.

Our results also contribute to the debate on the universality classes of the
DT and of the WV models in $d=2$~\cite{Siegert,chatraphorn,toroczkai,CTS}. For
both models, there is evidence of an asymptotic EW behavior. In the case of 
the DT model, the possibility of VLDS behavior is excluded by the evolution 
of the effective dynamical exponents. For  the  WV model, our 
result is in contradiction with the universality class suggested by Das Sarma and 
co-workers~\cite{chatraphorn,toroczkai,CTS}, 
the unstable(mounded morphology) class, but agrees with previous 
results of Siegert~\cite{Siegert} and Krug et al~\cite{KPS}.
It motivates further numerical studies on these lines, although the 
computational cost will significantly increase, mainly due to the rapid 
increase of fluctuations in the $w$ and in the $\xi$ data as the deposition 
time increases.

We note  that, if the asymptotic universality class of the
WV model in $d=2$ is in fact the unstable (mounded morphology) one  suggested
by Das Sarma and co-workers, it is not particularly meaningful to talk about
the usual exponents  $\alpha$, $\beta$ and $z$ anymore, since the surface would
not be statistically scale invariant. Possibly an explanation for the 
controversy could come from the observation of the morphologies of 
the growth fronts in the simulations~\cite{CTS} using noise reduction techniques. 
While the DT surface gets smoother when the noise 
reduction factor varies from 1 (no noise reduction) to 5 (Fig.4 in 
Ref. \protect\cite{CTS}),  the morphology of the WV surface
significantly changes, from an irregular surface with
no noise reduction to an organized mounded surface for noise reduction
factor $5$  (Fig. 5 in Ref. \protect\cite{CTS}). 
 
From the analytical point of view, some
progress is expected after the recent works of Vvedensky and
co-workers~\cite{baggio,vvedensky}, although the application of their methods
to systems in $d=2$ seems to be much harder. 

\vskip 1cm

{\bf Acknowledgements}

This work was partially supported by CNPq and FAPERJ (Brazilian agencies).


\end{document}